\documentclass[amsmath,twocolumn,aps]{revtex4}

\usepackage{longtable}
\usepackage{latexsym}
\usepackage{amssymb}
\usepackage{epsfig}
\usepackage{amsmath}
\usepackage{float}

\newcommand{\adotoa}{\ensuremath{{\cal H}}}
\newcommand{\GammaGI}{\ensuremath{\hat{\Gamma}}}
\newcommand{\PhiGI}{\ensuremath{\hat{\Phi}}}
\newcommand{\PsiGI}{\ensuremath{\hat{\Psi}}}
\newcommand{\deltaGI}{\ensuremath{\hat{\delta}}}
\newcommand{\thetaGI}{\ensuremath{\hat{\theta}}}
\newcommand{\PiGI}{\ensuremath{\hat{\Pi}}}
\newcommand{\half}{\ensuremath{\frac{1}{2}}}
\newcommand{\coupsign}{\ensuremath{s}}

\newcommand{\grad}{\ensuremath{\vec{\nabla}}}

\newcommand{\Jb}{\ensuremath{\bar{J}}}
\newcommand{\Pb}{\ensuremath{\bar{P}}}
\newcommand{\Tb}{\ensuremath{\bar{T}}}
\newcommand{\rhob}{\ensuremath{\bar{\rho}}}

\newcommand{\phib}{\ensuremath{\bar{\phi}}}
\newcommand{\Zb}{\ensuremath{\bar{Z}}}
\newcommand{\Fb}{\ensuremath{\bar{F}}}
\newcommand{\Kb}{\ensuremath{\bar{K}}}

\newcommand{\Xb}{\ensuremath{\bar{X}}}

\newcommand{\Lie}[1]{\ensuremath{\underset{#1}{\mathcal{L}}}}

\newcommand{\ParDiff}{\ensuremath{{\mu}}}

\newcommand{\be}{\begin{equation}}
\newcommand{\ee}{\end{equation}}
\newcommand{\bea}{\begin{eqnarray}}
\newcommand{\eea}{\end{eqnarray}}

\begin{document}

\title{The Parameterized Post-Friedmannian Framework for Interacting Dark Energy Theories}

\author{
C.~Skordis\footnote{E-mail:  skordis@ucy.ac.cy}$^{a}$,
A.~Pourtsidou\footnote{E-mail: alkistis.pourtsidou@port.ac.uk}$^{b}$,
E.~J.~Copeland\footnote{E-mail: ed.copeland@nottingham.ac.uk}$^{c}$
}

\affiliation{
$^a$ Department of Physics, University of Cyprus, 1 University Avenue, Nicosia 2109, Cyprus
\\
$^b$ Institute of Cosmology \& Gravitation, University of Portsmouth, Dennis Sciama Building, Burnaby Road, Portsmouth, PO1 3FX, United Kingdom
\\
$^c$School of Physics and Astronomy, University of Nottingham, University Park, Nottingham NG7 2RD, UK 
}

\begin{abstract}
We present the most general parameterization of models of dark energy in the form of a scalar field 
which is explicitly coupled to dark matter. We follow and extend the Parameterized Post-Friedmannian approach, previously applied to modified gravity theories, 
in order to include interacting dark energy. We demonstrate its use through a number of worked examples and show how the initially large parameter space of free functions can be significantly reduced and constrained to include only a few non-zero coefficients. This paves the way for a model-independent approach to classify and test interacting dark energy theories.
\end{abstract}

\maketitle

\section{Introduction}
In recent years, cosmological data from experiments with exquisite precision (Cosmic Microwave Background measurements (\cite{Komatsu:2010fb},\cite{Ade:2013zuv}), Ia supernovae~\cite{Kowalski:2008ez}, Baryon Acoustic Oscillation surveys~\cite{Lampeitl:2009jq}) suggest that $\sim 96\%$ of the matter/energy
content of our Universe is in the form of an exotic dark sector.  Approximately a quarter of the dark sector is believed to be weakly interacting Cold Dark Matter, while roughly $70\%$ 
is in the form of a Dark Energy component, a substance  with negative pressure responsible for the current accelerated expansion of the Universe. 

The best candidate for Dark Energy is the cosmological constant $\Lambda$. The concordance model of cosmology, $\Lambda$CDM, is currently the best fit to observations, but it comes along with fundamental questions and problems. One of them is the coincidence problem, which poses the question why the energy densities of the dark sector components are of the same order today, when their cosmological evolution is very different. A possible solution to the coincidence problem is a coupling between the Dark Energy and the Dark Matter. The introduction of an appropriate coupling does not violate observational constraints, and it can change the background evolution of the dark sector components in order to offer a solution to the coincidence problem. 

A plethora of such dark coupling models can be found in the literature (see, e.g. \cite{Amendola:1999er, Billyard:2000bh, Zimdahl:2001ar, Farrar:2003uw, Matarrese:2003tn, Amendola:2003wa, Maccio:2003yk, Amendola:2004ew, Amendola:2006dg, Guo:2007zk, Koivisto:2005nr, Lee:2006za, Wang:2006qw, Mainini:2007ft, Pettorino:2008ez, Xia:2009zzb, Chimento:2003iea, Olivares:2005tb, Sadjadi:2006qp, Brookfield:2007au, Boehmer:2008av, CalderaCabral:2009ja, He:2008tn, Quartin:2008px, Valiviita:2009nu, Pereira:2008at, Bean:2008ac, Gavela:2009cy, Simpson:2010vh, Tarrant:2011qe,Koivisto:2013fta,Salvatelli:2014zta,Boehmer:2015aa,Boehmer:2015sha}). In most of these models, the choice of the coupling is purely phenomenological. In a recent paper \cite{Pourtsidou:2013aa}, we made further progress at the level of construction of such models by identifying three separate $\emph{classes}$ of models of dark energy in the form of a scalar field $(\phi)$ coupled to Cold Dark Matter (CDM). 

After constructing general models of exotic dark energy or modified gravity and checking their mathematical and physical viability (for example by identifying fundamental problems
 like ghosts or strong coupling issues), one is interested in testing them against the available data to see if they might offer a viable alternative to $\Lambda$CDM. 

Currently, there is a pressing need of fast and efficient ways to rule out and constrain the large number of cosmological models available --- it would be practically impossible to go through each and every one of them individually. The Parameterized Post-Friedmannian (PPF) approach offers such a framework and has been applied to modified gravity theories ~\cite{Skordis:2008vt,Baker:2011jy,Baker:2012zs} 
 (see~\cite{Ferreira:2014mja} for a recent overview). In this work we apply the PPF approach in interacting dark energy theories and demonstrate its use through a number of worked examples. In Section~\ref{sec:formalism} we go through the PPF basic principles and general formalism, extending it to the case of coupled dark matter/dark energy. In Section~\ref{sec:examples} we first demonstrate how a few of the most well-known phenomenological models in the literature fit in to this formalism and then we proceed to apply it to the general classes of models we presented in \cite{Pourtsidou:2013aa}. We conclude in Section~\ref{sec:conclusions}.

\section{Formalism}
\label{sec:formalism}

\subsection{Basic Concepts}
We start by writing the gravitational field equations of a theory as
\be
G_{\mu \nu} =  8\pi G \left( T^{(SM)}_{\mu \nu}+ T^{(GDM)}_{\mu \nu} + T^{(DE)}_{\mu \nu}  \right),
\label{eq:einstein}
\ee 
where $G_{\mu\nu}$ is the Einstein tensor of the metric $g_{\mu\nu}$, $T^{(SM)}_{\mu\nu}$ is the stress-energy tensor
of the \emph{known} forms of matter (baryons, photons, neutrinos, etc) that are part of the Standard Model of particle physics,
 $T^{(GDM)}_{\mu \nu}$ is the stress-energy tensor of (Generalized) Dark Matter and $T^{(DE)}_{\mu \nu}$
represents the stress-energy tensor of all the unknown modifications to the gravitational field equations that generate the effect of Dark Energy.
 Such modifications may be purely due to a Dark Energy fluid or perhaps due to a modification of gravity. It may be shown that any kind
of modification of gravity can be put in the form (\ref{eq:einstein}) (see for instance ~\cite{Skordis:2008vt}). Let us also note 
that although we start with a Generalized Dark Matter which may have non-zero pressure and non-zero shear~\cite{Hu:1997},
we shall later on specialize to the Cold Dark Matter (CDM) case where both of these quantities are zero.

The Bianchi identities tell us that the Einstein tensor is divergenceless:
\be
\nabla_\mu G^\mu_{\;\;\nu}=0.
\ee
which in turn implies that 
 $\nabla_\mu  \left(  T^{(SM)\mu}_{\phantom{(SM)\mu} \nu} +  T^{(GDM)\mu}_{\phantom{(GDM)\mu} \nu} + T^{(DE)\mu}_{\phantom{(DE)\mu} \nu} \right) =0$.
We assume that the standard model particles do not explicitly couple to the dark sector so that $\nabla_\mu  T^{(SM)\mu}_{\phantom{(SM)\mu} \nu} =0$.
 This assumption is well justified by observations which strongly constraints such couplings \cite{Carroll:1998zi}.
 Furthermore, a coupling of the evolving quintessence field to baryons would lead to time varying constants of nature, 
which are tightly constrained, see \cite{Martins:2009zz} and references therein.
This leaves us with $\nabla_\mu  \left(T^{(GDM)\mu}_{\phantom{(GDM)\mu} \nu} + T^{(DE)\mu}_{\phantom{(DE)\mu} \nu} \right) =0$ 
but neither part is assumed to be individually conserved. Thus we have that
\be
  \nabla_\mu T^{(GDM)\mu}_{\phantom{(GDM)\mu} \nu}   = J_\nu = - \nabla_\mu  T^{(DE)\mu}_{\phantom{(DE)\mu}\nu}
\label{eq:Jcurrent}
\ee
 where the coupling current  $J_\nu$ represents the energy and momentum exchange between the dark sector components.

In what follows we aim to parametrise the coupling current  $J_\nu$  in terms of metric potentials and their derivatives 
as well as the scalar modes that are part of the stress-energy tensors of the two dark sector components. 
We shall do that in such a way so that the resulting field equations contain at most two time derivatives, or equivalently, 
each dark sector component obeys two first order linearized field equations on an FRW background resulting from (\ref{eq:Jcurrent}).
We shall proceed by considering first a FRW background spacetime and find the relevant equations that describe the dark sector
 and then consider linear perturbations about this background spacetime and see how this affects the parameterization. Background variables will 
be signified with a ``bar'' (unless no confusion can arise, e.g. the scale factor $a$ is always a background variable) 
while typically all perturbed tensors will be preceded by a $\delta$. For instance, we may split $J_\nu$ into 
$J_\nu = \Jb_\nu + \delta J_\nu$.

\subsection{FRW Background}
Consider a FRW background spacetime described by a metric
\be
  ds^2 = a^2\left(-d\tau^2 + \gamma_{ij} dx^i dx^j\right)
\label{FRW_metric}
\ee
where $a$ is the scale factor, $\tau$ is the conformal time and  $\gamma_{ij}$  is the spatial metric, assumed to be flat. 
The symmetries of the spacetime impose that the only non-zero components $\Tb_{\mu\nu}$ are the energy density
 $\rhob=-\Tb^0_{\;\;0}$ and pressure $\Pb$ such that $\Tb^i_{\phantom{i}j} = \Pb\delta^i_{\phantom{i}j}$.

The generalized Einstein equations (\ref{eq:einstein}) for this ansatz give
\be
3\adotoa^2 = 8\pi G a^2 \left( \rhob_{SM} + \rhob_{GDM} + \rhob_{DE} \right)
\ee and
\be
\adotoa^2  -2 \frac{\ddot{a}}{a}= 8\pi G \left( \Pb_{SM} + \Pb_{GDM} + \Pb_{DE} \right), 
\ee
where $\adotoa = \frac{\dot{a}}{a}$ is the conformal Hubble parameter and dots denote derivatives with respect to $\tau$. 

Turning now to the coupling current $J_\nu$, the symmetries of the spacetime impose that the only non-zero component is
\be
Q \equiv \Jb_0
\ee
while $\Jb_i=0$. The function $Q(\tau)$ is the background coupling function which is for our purposes a phenomenologically free function. 
Specific models of a coupled dark sector will in general result to specific choices of $Q(\tau)$  
(see, for example, \cite{Amendola:1999er, Boehmer:2008av}, which are two models we present and parametrise in Section~\ref{sec:examples}).

The $\nu=0$  component of (\ref{eq:Jcurrent}) gives the field equations for the evolution of a particular  component indexed by ``$I$'' as
\be
\dot{\rhob}_{I} +3 \adotoa \rhob_{I} (1 + w_{I} )= s_I Q,
\label{dot_rho}
\ee
 where we have defined the equation of state parameter for each $I$-component as $w_I \equiv \Pb_I/\rhob_I$ and
the constant $\coupsign_I$ takes the values 
\be
 \coupsign_I = \bigg\{\begin{matrix}  1 \qquad \qquad \mbox{DE} \\ \quad \quad   \; \;  0  \qquad \qquad  \mbox{SM fields} \\  \, \; \;  -1 \qquad \qquad  \mbox{GDM} \end{matrix}
.
\ee

\subsection{Linear Perturbations}

\subsubsection{The perturbed variables}
We now turn to linear perturbations about the FRW background. We shall consider only scalar modes. The spacetime metric takes the form
\bea 
\nonumber
ds^2&=-a^2(1+2\Psi)dt^2 - 2a^2\nabla_i \zeta  dtdx^i \\
&+ \; a^2\left[(1+\frac{1}{3}h) \gamma_{ij} + D_{ij}\nu \right]dx^idx^j,
\label{perturbed_metric}
\eea 
where $\Psi$, $\zeta$, $h$, and $\nu$ are four functions of time and space (four scalar modes) and
\be
D_{ij} = \nabla_i \nabla_j -\frac{1}{3}\gamma_{ij}\nabla^2
\ee
 is a derivative operator that projects out the longitudinal, traceless, spatial part of the perturbation. 

Let us now consider the perturbed variables of the fluids. These are the density contrast $\delta \equiv \delta \rho /\rhob$, 
the scalar mode of the momentum, $\theta$, such that $u_i=a\nabla_i \theta$, the dimensionless pressure perturbation 
$\Pi \equiv \delta P/\rhob$ such that $\delta T^i_j = \Pi \rhob \delta^i_{\;j}$ and the scalar mode of the shear $\Sigma$ 
such that the shear tensor is
$\Sigma_{ij} = D_{ij}\Sigma$. Putting it all together, the  stress-energy tensor components for a fluid are
\begin{eqnarray}
T^0_{\phantom{0}0} &=& -\rhob (1 + \delta)
\\
T^0_{\phantom{0}i} &=& -(\rhob + \Pb)  \grad_i \theta 
\\
T^i_{\phantom{i}0} &=& (\rhob + \Pb)  \grad^i(\theta-\zeta)  
\\
T^i_{\phantom{i}j} &=&  \rhob (w  + \Pi) \delta^i_{\phantom{i}j} +  (\rhob +\Pb)  D^i_{\phantom{i}j} \Sigma
\end{eqnarray}

\subsubsection{Einstein and fluid equations}

The perturbed Einstein equations (\ref{eq:einstein}) are
\begin{subequations} 
\begin{equation}
  \adotoa \left(\dot{h} + 2  \grad^2  \zeta  \right) -6\adotoa^2   \Psi  + 2 \grad^2 \eta = 8\pi G a^2 \sum_I \rhob_I \delta_I,
\end{equation}
\begin{equation}
 2 \dot{\eta}  + 2  \adotoa \Psi   =  8\pi G a^2 \sum_I (\rhob_I + \Pb_I)  \theta_I,
\end{equation}
\begin{eqnarray}
&&
-  \ddot{h} -   2 \adotoa \dot{h} + 6 \adotoa \dot{\Psi} + 6 \left( \adotoa^2  +2\dot{\adotoa}   \right) \Psi 
\nonumber 
\\
&&
\qquad \qquad
-   \grad^2 \left(  2 \eta - 2 \Psi   + 2 \dot{\zeta} +  4 \adotoa \zeta \right)
\nonumber 
\\
&&
\qquad \qquad
= 24 \pi G a^2 \sum_I \rhob_I \Pi_I
\end{eqnarray}
and
\begin{eqnarray}
&&
 \frac{1}{2}   \ddot{\nu} + \dot{\zeta} + \adotoa \left(\dot{\nu} + 2  \zeta \right)  +  \eta -\Psi 
\nonumber 
\\
&&
\qquad \qquad
= 8\pi G a^2\sum_I (\rhob_I +\Pb_I) \Sigma_I
\end{eqnarray}
\label{perturbed_einstein}
\end{subequations}

Turning now to the fluid equations, they are obtained by perturbing (\ref{eq:Jcurrent}). 
To this purpose we define the two scalar mode perturbations $q$ and $S$ by
\be 
 q \equiv \delta J_0 
\qquad 
\qquad 
 \grad_i S \equiv \delta J_i  
. 
\ee
We find
\begin{subequations} 
\bea 
\dot{\delta}_I &=& 3w_I \adotoa \delta_I 
 + (1+w_I)\left[ \grad^2\theta_I -\half  \dot{h}-\grad^2 \zeta \right]
\nonumber
\\
&&
-3\adotoa\Pi_I
+ \frac{\coupsign_I}{\rhob_I} \left[q - Q\delta_I  \right]
, 
\eea
and
\bea
\dot{\theta}_I &=&
-\left[ \adotoa(1-3w_I)+\frac{\dot{w}_I}{1+w_I}\right]\theta_I +\frac{\Pi_I}{1+w_I} 
 \nonumber
\\
&&
+\frac{2}{3}\grad^2\Sigma_I
+\Psi 
+ \frac{\coupsign_I}{\rhob_I} \left[\frac{S}{1+w_I} - Q \theta_I \right]
\, .
\eea
\label{perturbed_fuid}
\end{subequations} 
where the index $I$ runs over all species (and once again let us recall that $s_{DE} = 1 = -s_{GDM} $ while $s_I =0$ for all other species).

\subsection{Dark Coupling Parameterisation}
The goal of this article is to parametrise both of the two perturbation variables $q$ and $S$
 as linear combinations of all other perturbations, such as, the fluid variables $\delta$, $\theta$, $\Pi$ and $\Sigma$ for each fluid, as well as the metric
variables $\Psi$, $\zeta$, $h$ and $\nu$. This means 12 variables in total for each of $q$ and $S$. 
However, this linear combination is not entirely arbitrary, but must obey certain rules regarding gauge transformations. As we shall see, this reduces the number of
effective independent variables to $10$ for each of  $q$ and $S$. 
Before proceeding to the parameterization, let us briefly discuss gauge transformations.

\subsubsection{Gauge transformations}
The metric in (\ref{perturbed_metric}) is in a form which is not gauge-fixed.  In other words the four scalar modes are not invariant
under gauge transformations $\delta g_{\mu\nu} \rightarrow \delta g_{\mu\nu} + \Lie{\xi} \bar{g}_{\mu\nu} $ generated by a vector field $\xi^\mu$.  Parameterized as $\xi^\mu = \frac{1}{a} \left(\xi_T, \grad^i \xi_L\right)$
for two scalar  modes $\xi_T$ and $\xi_L$ the gauge transformations of the metric and fluid perturbations will involve combinations of  $\xi_T$ and 
$\xi_L$ and their first time derivatives. 

Consider first the variables $q$ and $S$. We find that they transform as
\begin{subequations}
\be
q \rightarrow q + \frac{1}{a}\left[ Q \dot{\xi}_T  + \left( \dot{Q}  - \adotoa Q \right) \xi_T\right]
\ee
and
\be
S \rightarrow S + \frac{1}{a}Q\xi_T,
\ee
respectively.
\end{subequations}
Thus if we write $q$ and $S$ as a linear combination of the metric and fluid variables, variables which involve $\xi_L$ in their transformation 
must combine together so that $\xi_L$ does not appear overall in the transformation of the entire linear combination.

Now the fluid variables transform only with the gauge variable $\xi_T$, i.e. as (dropping the obvious $I$ indices)
\begin{subequations}
\bea 
 \delta   &\rightarrow&   \delta   - \frac{1}{a} \left[ 3 \adotoa (1 + w )  -  \coupsign \frac{Q}{\rhob} \right] \xi_T,
 \\ 
 \theta &\rightarrow&  \theta + \frac{1}{a} \xi_T, \\ 
\Pi   &\rightarrow&   \Pi  + \frac{1}{a}\left[  \dot{w} -3\adotoa (1 + w) w + \coupsign w \frac{Q}{\rhob} \right] \xi_T ,
\eea 
\end{subequations}
while $\Sigma$ is gauge-invariant, hence, all four of them are allowed to appear in the $q$ and $S$ parameterization.

However, the metric variables involve $\xi_L$ in their transformation. This means that the metric variables must combine together so that $\xi_L$ is 
eliminated alltogether.  Following  \cite{Skordis:2008vt} we can find three linear combinations of the metric perturbations and their first time derivatives
 which transform only with the gauge variable $\xi_T$.
These are $U \equiv h  -\nabla^2 \nu$ and  $V\equiv \dot{\nu} + 2\zeta$ as well as
$\dot{h}+2\nabla^2 \zeta$.  The latter one is not independent but is equal to $\dot{U} + \nabla^2 V$. Thus out of the four metric scalar modes, we are left
with two combinations, namely $U$ and $V$, which transform exclusively with $\xi_T$
and $\Psi$ which transforms with $\dot{\xi}_T$ . Explicitely, the transformations are
\begin{subequations}
\bea 
 U   &\rightarrow&   U   + \frac{6}{a}\adotoa \xi_T,
 \\ 
 V &\rightarrow&  V  + \frac{2}{a} \xi_T, \\
\Psi   &\rightarrow&    \Psi  + \frac{\dot{\xi}_T}{a}\, .
\eea
\end{subequations}
Since,  $q$ and $\Psi$ contain $\dot{\xi}_T$ in their transformation, we must allow a further metric variable combination which does so, but
which doesn't have higher than 2nd time derivatives.  The only possibility is the variable $\dot{U}$.

To summarise, we expect that each of $q$ and $S$ can be written as linear combinations of the four fluid variables (for each fluid) plus the four
variables $U$, $V$, $\Psi$ and $\dot{U}$.

\subsubsection{Completing the parameterization}
Following the discussion above, we start from the parameterization
\bea
q &=&   C_1 \Psi + C_2 V  + A_1 U + A_2 \dot{U} 
 +  A_3 \delta_{DE}  +  A_4 \delta_{GDM}
 \nonumber
\\ 
&&
 + A_5 \theta_{DE}   + A_6 \theta_{GDM} +   A_7 \Pi_{DE}  +  A_8 \Pi_{GDM}
\nonumber 
\\
&&
  + A_9 \Sigma_{DE} + A_{10} \Sigma_{GDM}
\label{q_param_basic}
\eea
and
\bea
S &=&  C_3 \Psi + C_4 V  + B_1 U + B_2 \dot{U} 
 +  B_3 \delta_{DE}  +  B_4 \delta_{GDM}
 \nonumber
\\ 
&&
 + B_5 \theta_{DE}   + B_6 \theta_{GDM} +   B_7 \Pi_{DE}  +  B_8 \Pi_{GDM}
\nonumber 
\\
&&
  + B_9 \Sigma_{DE} + B_{10} \Sigma_{GDM}
\label{S_param_basic}
\eea
Performing the gauge transformations in  (\ref{q_param_basic})  we find two constraint equations, namely
\begin{subequations}
\be
C_1  = Q - 6 \adotoa A_2
\label{eq:constr1}
\ee
 and
\bea
 2  C_2  &=& \dot{Q}  - \adotoa Q - 6 \adotoa A_1 +6 ( \adotoa^2  - \dot{\adotoa} ) A_2 
- \frac{\dot{\rhob}_{DE}}{\rhob_{DE}}A_3
 \nonumber
\\ 
&&
 -  \frac{\dot{\rhob}_{GDM}}{\rhob_{GDM}} A_4 - A_5 - A_6  
 - \frac{ \dot{\Pb}_{DE} }{\rhob_{DE}} A_7
 - \frac{ \dot{\Pb}_{GDM} }{\rhob_{GDM}} A_8
\qquad
\label{eq:constr2}
\eea
Likewise, performing the gauge transformations in (\ref{S_param_basic}) we find two further constraint equations, namely
\be
C_3 =  - 6 B_2\adotoa 
\label{eq:constr3}
\ee
and
\bea
  2 C_4 
&=&
Q-  6 \adotoa B_1
+ 6(\adotoa^2 - \dot{\adotoa}) B_2
 -  \frac{\dot{\rhob}_{DE}}{\rhob_{DE}}  B_3
 \nonumber
\\ 
&&
 -  \frac{\dot{\rhob}_{GDM}}{\rhob_{GDM}}  B_4
 - B_5    - B_6 
 - \frac{\dot{\Pb}_{DE}}{\rhob_{DE}}  B_7 
 - \frac{\dot{\Pb}_{GDM}}{\rhob_{GDM}}    B_8 \;\;
\qquad
\label{eq:constr4}
\eea
\end{subequations}
The  two constraints (\ref{eq:constr1}) and (\ref{eq:constr2}) are then used to eliminate $C_1$ and $C_2$ from (\ref{q_param_basic}) while 
the  two constraints (\ref{eq:constr3}) and (\ref{eq:constr4}) are used to eliminate $C_3$ and $C_4$ from (\ref{S_param_basic}).
The remaining perturbations are written in terms of the gauge-invariant variables listed in Table~\ref{tab:GI}
by combining them with $V$. The result is
\bea
q &=& \half (\dot{Q}  - \adotoa Q) V + Q \Psi  - 6 A_1 \PhiGI - 6 A_2 \GammaGI 
 \nonumber
\\ 
&&
 +  A_3 \deltaGI_{DE} +  A_4 \deltaGI_{GDM} + A_5 \thetaGI_{DE}   + A_6 \thetaGI_{GDM}
 \nonumber
\\ 
&&
 +   A_7 \PiGI_{DE}  +  A_8 \PiGI_{GDM}
  + A_9 \Sigma_{DE} + A_{10} \Sigma_{GDM}
\qquad
\label{q_param_GI}
\eea
and
\bea
S &=&\half Q V - 6 B_1 \PhiGI  - 6 B_2 \GammaGI +  B_3 \deltaGI_{DE}  +  B_4 \deltaGI_{GDM}
 \nonumber
\\ 
&&
 + B_5 \thetaGI_{DE}   + B_6 \thetaGI_{GDM} +   B_7 \PiGI_{DE}  +  B_8 \PiGI_{GDM}
\nonumber 
\\
&&
  + B_9 \Sigma_{DE} + B_{10} \Sigma_{GDM}
\label{S_param_GI}
\eea
Hence, we are left with $20$ free functions in total.

\begin{table}
\centering
\caption{Gauge invariant variables.
\label{tab:GI}}
\small
  \begin{tabular}{l}
\hline
\\
$\PhiGI \equiv  -\frac{1}{6} U + \half \adotoa V$
\\
$\PsiGI \equiv \Psi -\frac{1}{2}\dot{V}-\frac{1}{2}\adotoa V$
\\
$\GammaGI \equiv -\frac{1}{6} \dot{U} +  \adotoa \Psi + \half(\dot{\adotoa} -  \adotoa^2 )V =  \dot{\hat{\Phi}}+ \adotoa \hat{\Psi} $
\\
$\deltaGI \equiv \delta -  \half \frac{\dot{\rhob}}{\rhob} V $
\\
$ \thetaGI \equiv \theta - \frac{1}{2} V$
\\
 $\PiGI \equiv \Pi -  \frac{ \dot{\Pb}}{\rhob}V$
\\ \\
\hline
\hline
 \end{tabular}
\end{table}

\subsubsection{Special case: Cold Dark Matter}
From now on we will assume that the Dark Matter fluid is completely cold. This automatically means that $w_{GDM}= \Pi_{GDM} = \Sigma_{GDM} = 0$.
We shall further make the assumption that  the Dark Energy fluid has no shear, i.e. $\Sigma_{DE} = 0$. Furthermore, since there is no possibility of confusion
we shall set $w_{DE} = w$.

In general, the pressure perturbation $\Pi_{DE}$ would be an independent dynamical degree of freedom (see~\cite{BanadosFerreiraSkordis2009} for an explicit model).
However, there are many instances where  $\Pi_{DE}$ is expressed in terms of $\delta_{DE}$ and $\theta_{DE}$ via equations of state such as
 the Generalized Dark Matter model~\cite{Hu:1997}. 
As in ~\cite{Hu:1997} we shall also assume that  the pressure perturbation $\Pi_{DE}$ is  expressed in terms of $\delta_{DE}$ and $\theta_{DE}$ 
via equations of state. However, the usual expression in ~\cite{Hu:1997} no longer holds, 
as it does not transform correctly  under gauge transformations. An expression which does is 
\begin{eqnarray}
\Pi_{DE} &=& c^2_s \delta_{DE} +(c^2_s-c^2_a) \left[ 3(1+w)\adotoa -\frac{Q}{\rhob_{DE}}\right] \theta_{DE}
\nonumber
\\
&&
\ \ \ \
 + \ParDiff (\theta_c - \theta_{DE})
\label{eq:PiXcoupled}
\end{eqnarray}
where $c_s^2$ and $c_a^2$  are the (gauge-invariant) effective and adiabatic speeds of sound respectively.
It may be shown that the divergence of the entropy flux is proportional to $\Pi_{DE} - c_a^2\delta_{DE}$~\cite{DurrerBook}, hence,
the gauge-invariant ``relative entropy'' parameter $\mu$ measures entropy transfer to DE due to its motion relative to the CDM fluid.
 The adiabatic speed of sound is fixed by the equation of state $w$ via
\be
c^2_{a} = w+\frac{\dot{w}}{\frac{Q}{\rhob_{DE}}-3\adotoa (1+w)}.
\label{eq:c2a}
\ee
Hence, without loss of generality, we may further set $A_7$ and $B_7$ to zero.  With these choices, the number of free functions
is reduced to $12$.

We shall further assume the conformal Newtonian gauge for which $\zeta = \nu = 0$ (so that $V = 0$). With this choice,
 the gauge-invariant variables we have defined in table \ref{tab:GI} are equal to the conformal Newtonian gauge variables.

Let us now re-state the parameterization as well as the necessary evolution equations. The two parameters $q$ and $S$ are given by
\begin{subequations}
\bea
q &=& Q \Psi  - 6 A_1 \Phi - 6 A_2 (\dot{\Phi} + \adotoa\Psi) +  A_3 \delta_{DE} 
 \nonumber
\\ 
&&
+  A_4 \delta_c + A_5 \theta_{DE}   + A_6 \theta_c
\qquad
\label{q_param_CN}
\eea
and
\bea
S &=& - 6 B_1 \Phi  - 6 B_2(\dot{\Phi} + \adotoa \Psi)  +  B_3 \delta_{DE}  
 \nonumber
\\ 
&&
+  B_4 \delta_c + B_5 \theta_{DE}   + B_6 \theta_c
\label{S_param_CN}
\eea
\end{subequations}
for unknown functions $A_i$ and $B_i$ with $i \in 1\ldots 6$. 

The evolution equations for CDM are
\begin{subequations} 
\bea 
\dot{\delta}_c &=&  \grad^2\theta_c + 3 \dot{\Phi}  + \frac{1}{\rhob_c} \left(Q\delta_c - q \right) , 
\eea
and
\bea
\dot{\theta}_c &=& - \adotoa\theta_c +\Psi + \frac{1}{\rhob_c} \left( Q \theta_c - S \right) \, .
\eea
\label{CDM_eq_CN}
\end{subequations} 
while the evolution equations for DE are
\begin{subequations} 
\bea 
\dot{\delta}_{DE} &=& 3w \adotoa \delta_{DE} 
 + (1+w)\left[ \grad^2\theta_{DE} + 3  \dot{\Phi} \right]
\nonumber
\\
&&
-3\adotoa\Pi_{DE}
+ \frac{1}{\rhob_{DE}} \left[q- Q\delta_{DE} \right]
, 
\eea
and
\bea
\dot{\theta}_{DE} &=&
-\left[ \adotoa(1-3w)+\frac{\dot{w}}{1+w}\right]\theta_{DE} +\frac{\Pi_{DE}}{1+w} 
 \nonumber
\\
&&
+\Psi 
+ \frac{1}{\rhob_{DE}} \left[\frac{S}{1+w} - Q \theta_{DE} \right]
\, .
\eea
\label{DE_eq_CN}
\end{subequations} 

 In the following Section we are going to investigate the underlying space
of models of coupled DM to DE, and show how we can construct a ``dictionary" of interacting dark energy theories and their PPF correspondences. The same method was applied to modified gravity theories in \cite{Baker:2012zs}.

\section{Worked Examples}
\label{sec:examples}

As a ``warm-up" exercise, we are first going to demonstrate the use of our PPF formalism for interacting dark energy theories by showing that the functions $A_i$ and $B_i$ are severely constrained when one considers specific models which appear often in the literature. These are  the ``coupled quintessence'' model \cite{Amendola:1999er}, 
a model where $J_\mu \propto u_\mu$ \cite{Boehmer:2008av,Valiviita:2008iv}
and the elastic scattering of model of Dark Matter and Dark Energy \cite{Simpson:2010vh,Baldi:2014aa}. In table \ref{tab_of_models} one can see the list of the models we consider with their coefficients displayed.

Following that, we consider the parameterisation of the general \emph{classes} of coupled theories we constructed in \cite{Pourtsidou:2013aa}. More specifically, in \cite{Pourtsidou:2013aa} we presented three distinct types of models of dark energy in the form of a scalar field explicitly coupled to dark matter. 
We used the pull-back formalism for fluids and generalized the standard fluid action in order to include a dark coupling. The general functional form for the combined 
dark energy and dark matter Lagrangian we considered is
\be
L=L(n,Y,Z,\phi),
\label{gen_L}
\ee where $n$ is the fluid number density, $Y=\half \nabla_\mu \phi \nabla^\mu \phi$, and $Z=u^\mu \nabla_\mu \phi$. 
As an example, within GR, a quintessence field and an uncoupled fluid is described by the Lagrangian $L = Y + V(\phi) + f(n)$.

 We then considered three distinct ways to reduce the general function (\ref{gen_L}) giving rise to the three Types of coupled models which we now want to  parametrise. These are the 
Type-1 models where $L = F(Y,\phi) + f(n,\phi)$ (the coupled quintessence model \cite{Amendola:1999er} is a subcase of Type-1 with the choice $F= Y + V(\phi)$ and $f= n e^{\beta_A \phi}$),
the Type-2 models where $L = F(Y,\phi) + f(n,Z)$ and the Type-3 models where  $L = F(Y,Z,\phi) + f(n)$.

\subsection{Specific models}
\subsubsection{Coupled Quintessence}
Let us start with the coupled quintessence (CQ) model suggested by Amendola \cite{Amendola:1999er}, which is a specific subcase of the Type 1 class of models we presented in  \cite{Pourtsidou:2013aa}. The scalar field action for this model is 
\be
S = -\int d^4x \sqrt{-g} \left[\half g^{\mu \nu}\partial_\mu \phi \partial_\nu \phi +V(\phi)\right],
\ee where $V(\phi)$ is the quintessence potential.
If a constant coupling parameter $\beta_A$ is assumed, the coupling current $J_\mu$ is found to be (see \cite{Pourtsidou:2013aa} for details)
\begin{equation}
  J_\mu = -\beta_A \rho_c \nabla_\mu \phi 
\label{coupled_quintessence_J}
\end{equation}
Writing the scalar field as $\phi=\phib + \varphi$ for a background field $\phib$ and perturbation $\varphi$,
 the components of the stress-energy tensor for this model are (using expressions from  \cite{Pourtsidou:2013aa})  
\bea
\rhob_{DE} = \frac{1}{2a^2}\dot{\phib}^2 + V  \qquad&&  \Pb_{DE} = \frac{1}{2a^2}\dot{\phib}^2 - V 
\\
c_a^2 = 1 + \frac{2\dot{\phib} V_\phi }{3\frac{\phib^2}{a^2}\adotoa - Q}
\qquad 
&&
Q = - \beta_A \rhob_c \dot{\phib} 
\eea
for the background, where $V_\phi \equiv \frac{dV}{d\phi}$ \, , and
\bea 
 \delta \rho_{DE} =\frac{\dot{\phib}}{a^2}(\dot{\varphi}-\dot{\phib}\Psi) +V_\phi \varphi \, ,
\quad &&
 \theta_{DE} =\frac{\varphi}{\dot{\phib}} \, ,
 \\
 \delta P_{DE} =\frac{\dot{\phib}}{a^2}(\dot{\varphi}-\dot{\phib}\Psi) -V_\phi \varphi \, ,
\quad &&
c_s^2 = 1\, , \quad \mu = 0
\quad
\eea
for the perturbations.
The required coupling parameters are found to be
\bea 
q &=& Q \left(\delta_c + \frac{\dot{\varphi}}{\dot{\phib}}\right) 
\label{eq:qCQ}
\\
 S&=&Q \frac{\varphi}{\dot{\phib}}.
\label{eq:SCQ}
\eea
Now we read-off the coefficients. They are:
\begin{eqnarray}
&& 
A_1 = A_2 = A_6 = 0 \qquad A_3 =  \frac{Q}{1 + w} 
\nonumber
\\
&&
  A_4 = Q \qquad
A_5 =\beta_A \rhob_c a^2 V_\phi 
\\
 && B_5 = Q \qquad  B_{i\ne 5} = 0
\nonumber
\end{eqnarray}

\subsubsection{Model with $J_\mu \propto u_\mu$}
In this model, which was introduced in \cite{Boehmer:2008av} and \cite{Valiviita:2008iv}, the energy-momentum transfer vector 
$J^\mu$ is parallel to the dark matter $4$-velocity $u^\mu$. In our notation we have 
$
u_\mu = a(1+\Psi, \nabla \theta_c)
$ and 
\be
J_\mu = \Gamma_{\rm int} \rhob_c (1+\delta_c)u_\mu,
\ee with $\Gamma_{\rm int}$ a local constant interaction rate. The  background  coupling function is 
\be
Q = a \Gamma_{\rm int} \rhob_c
\ee 
while the perturbative coupling parameters $q$ and $S$ are
\be
q= Q (\delta_c + \Psi) \qquad \mbox{and} \qquad  S=Q \theta_c.
\ee
Comparing with our general parameterisation scheme, we find that the only non-zero coefficients are 
\be
A_4=B_6= Q
\ee

\subsubsection{Elastic scattering of Dark Matter and Dark Energy}
\label{elastic_scattering_model}
This model was introduced in \cite{Simpson:2010vh} and it considered an elastic interaction between dark energy and dark matter. It is a pure momentum transfer model and its background cosmology remains unaltered. In our language, this model has 
\be
Q = 0
\ee in the background while
\be
q = 0 \qquad S = (\rhob_{DE} + \Pb_{DE})an_D\sigma_D (\theta_c - \theta_{DE}), 
\ee  at the level of the perturbations, with $n_D$ the proper number density of dark matter particles and $\sigma_D$ the scattering cross section between dark matter and dark energy (also note that $w={\rm const}$ and $c^2_{s} = 1$ in this model) \cite{Simpson:2010vh}.
 We therefore find that the only non-zero coefficients are
\be
B_5 =  - (\rhob_{DE} + \Pb_{DE})an_D\sigma_D  = -B_6.
\ee Now we turn our attention to the three general Types of models in \cite{Pourtsidou:2013aa}.

\subsection{Type 1 theory of DM coupled to DE}
Type 1 models are classified in  \cite{Pourtsidou:2013aa} via 
\be
L(n,Y,Z,\phi)= F(Y,\phi)+f(n,\phi).
\ee 
For the case where the Dark Matter is CDM we further have  $f(n,\phi)= n e^{\alpha(\phi)}$ where $\alpha(\phi)$ is a free function of the field $\phi$.

From \cite{Pourtsidou:2013aa} we have that the coupling current is 
\begin{equation}
J_\mu = - \rho_c \alpha_\phi \nabla_\mu \phi
\end{equation}
so that
\bea
Q &=& -\rhob_c \alpha_\phi \dot{\phib}
\nonumber
\\
q &=& Q\left( \delta_c + \frac{\dot{\varphi}}{\dot{\phib}}\right) - \rhob_c \alpha_{\phi\phi} \varphi
\label{eq:qSType1}
\\
S &=& Q \frac{\varphi}{\dot{\phib}}
\nonumber
\eea
where $\alpha_\phi \equiv \frac{d\alpha}{d\phi}$ and $\alpha_{\phi\phi} \equiv \frac{d^2\alpha}{d\phi^2}$.
Note that the expression for $S$ agrees with the one we recovered previously, Eq.~(\ref{eq:SCQ}), and the expression for $q$ agrees with Eq.~(\ref{eq:qCQ}) if $\alpha_{\phi\phi}=0$.
 This is expected, as the CQ model we studied is a sub case of Type 1.

We now need to express $\varphi$ and $\dot{\varphi}$ in terms of fluid-type variables. For Type-1 theories, we find from  \cite{Pourtsidou:2013aa} that
the background energy density and pressure are given by
\begin{equation}
\rhob_{DE} = -\Kb
\qquad
\Pb_{DE} = -\Fb
\label{eq:rhobPb}
\end{equation}
where we have introduced the function 
\begin{equation}
K(Y,\phi) = 2YF_Y - F
\label{K_def}
\end{equation}
which will come in handy below (and also for Type-2). Furthermore, the perturbed variables of interest are~\cite{Pourtsidou:2013aa}. 
\begin{align}
\delta_{DE} &= -\frac{\Zb \Kb_Y}{\Kb}  \delta Z + \frac{\Kb_\phi}{\Kb} \varphi \, ,  \qquad  \ \, \theta_{DE} =\frac{\varphi}{\dot{\phib}} \, ,
\\
c^2_a &= c_s^2 +\frac{\dot{\phib}\left[ \Fb_\phi - c_s^2 \Kb_\phi \right] }{ 3(\rhob_{DE} +\Pb_{DE})\adotoa -Q}  \, ,
\\
c_s^2 &= \frac{\Fb_Y}{\Kb_Y}  \, ,
 \qquad \qquad
 \qquad \qquad
 \qquad 
 \mu = 0
\label{eq:drhophidpphi}
\end{align}
where $\Kb_\phi \equiv \frac{\partial \Kb}{\partial \phi}$ and similarly for $\Fb$. 
The expression for $c_a^2$, above (and since also $\mu=0$),  says that 
if  $\Fb_\phi=0$  then the scalar field perturbations are adiabatic. 

To express  $\varphi$ and $\dot{\varphi}$ in terms of fluid-type variables we invert the relations (\ref{eq:drhophidpphi}) to get
\begin{equation}
\delta Z =   -  \frac{\Kb}{\Zb \Kb_Y} \delta_{DE} - \frac{a \Kb_\phi}{\Kb_Y} \theta_{DE}
\quad
,
\quad
\varphi = \dot{\phib} \theta_{DE}
\label{eq:deltaZgen}
\end{equation}
which are valid for adiabatic and non-adiabatic perturbations.

Let us now calculate the sought-after coefficients.  Using $a\delta Z=\dot{\phib}\Psi-\dot{\varphi}$ and (\ref{eq:deltaZgen})
 we find
\be
\frac{\dot{\varphi}}{\dot{\phib}}= \Psi  +  \frac{c_s^2}{1+w} \delta_{DE}
 - \frac{ c_s^2 \Kb_\phi}{(1+w)K} \dot{\phib} \, \theta_{DE}.
\ee
so that (\ref{eq:qSType1}) gives the required coefficients as
\begin{eqnarray}
&& 
A_1 = A_2 = A_6 = 0 \qquad A_3 =  \frac{Q c_s^2 }{1 + w} 
\nonumber
\\
&&
  A_4 = Q \qquad
A_5 =Q\left[\frac{\alpha_{\phi\phi}}{\alpha_\phi} - \frac{ c_s^2 \Kb_\phi}{(1+w) \Kb} \dot{\phib} \right]
\\
 && B_5 = Q \qquad  B_{i\ne 5} = 0
\nonumber
\end{eqnarray}
For the case where $\alpha_\phi = \beta_A $ is a constant and furthermore $F(Y,\phi) = Y + V(\phi)$ we recover the coefficients for the coupled quintessence
model discussed  above, which is in fact a subcase of a Type-1 model of coupled dark energy.

%%%%%%%%%%%%%%%%%%%%%%%%%%%%%%%%%%%%%%%%%%%%%%%%%%%%%%%%%%%%%%%%%%%%%%%%%%%%%%%%%%%%%%%%

\subsection{Type 2 theory of DM coupled to DE}

Type 2 models are classified via \cite{Pourtsidou:2013aa}
\be
L(n,Y,Z,\phi)=F(Y,\phi)+f(n,Z),
\ee with $f=nh(Z)$ in the case that the scalar field is coupled to CDM. 
The coupling current in this case is  \cite{Pourtsidou:2013aa}
\begin{equation}
 J_\mu = \nabla_\nu\left(\rho_c \beta u^\nu\right) \nabla_\mu\phi
\label{type_2_current}
\end{equation}
where $\beta(Z)$ is the function 
\be
\beta(Z) = \frac{h_Z}{h-Zh_Z}. 
\ee 
and $h_Z = \frac{dh}{dZ}$ (and the same when $Z$ is used as a subscript for $\beta$).
Let us first note that the relations for the fluid variables given by (\ref{eq:drhophidpphi}) are still valid for the case of Type-2 theory 
and the function $K$ is still defined via (\ref{K_def}).

In order to proceed further we need the function $Q$ which at first glance using (\ref{type_2_current}) is given by 
\begin{equation}
Q = \Zb \left[ \left(\dot{\rhob}_c + 3 \adotoa \rhob_c\right)\beta
 + \rhob_c \beta_Z \dot{\Zb} \right].
\label{Q_Z_dot_in}
\end{equation}
 Using (\ref{dot_rho}) to eliminate the terms $\dot{\rhob}_c$ and $\dot{\Zb}$ we find
\begin{equation}
Q = \frac{\Zb \beta_Z}{1+\Zb \beta} \rhob_c \dot{\Zb} 
\label{Q_Z_dot}
\end{equation}
where $\dot{\Zb}$ is determined from eq (70) of \cite{Pourtsidou:2013aa} as
\begin{equation}
\dot{\Zb} = -\frac{3 \Zb \Fb_Y  \adotoa  + a \Kb_\phi }{ \Kb_Y - \frac{\rhob_c \beta_Z}{1 + \Zb \beta}}
\end{equation}

The perturbative variables $q$ and $S$ are also found from (\ref{type_2_current}). Firstly $S$ is easily calculated as
\begin{equation}
S =  Q\theta_\phi
\end{equation}
while $q$ is found to be from (\ref{type_2_current}) as
\begin{eqnarray}
q &=& Q \Psi 
 + \Zb \bigg\{ \left(\dot{\delta}_c - \grad^2 \theta_c - 3 \dot{\Phi}\right) \rhob_c \beta
+ \rhob_c \beta_Z \dot{\delta Z}
\nonumber 
\\
&&
+ \left[ \left(\dot{\rhob}_c + 3\adotoa\rhob_c\right)\beta + \rhob_c \beta_Z \dot{\Zb}\right] \left[\delta_c - \Psi\right]
\nonumber 
\\
&&
+ \left[ \left(\dot{\rhob}_c + 3\adotoa\rhob_c\right)\beta_Z + \rhob_c \beta_{ZZ} \dot{\Zb}  +  \frac{Q}{\Zb^2} \right] \delta Z
\bigg\}
\end{eqnarray}
However, using (\ref{Q_Z_dot_in}) as well as (\ref{CDM_eq_CN}) in order to eliminate $\dot{\delta}_c$ the expression for $q$ simplifies to
\begin{equation}
 q =  Q \delta_c + Q \frac{\dot{\delta Z}}{\dot{\Zb}}
+ \frac{d}{dZ}\left[  \frac{  \Zb \beta_Z  }{1 + \Zb \beta} \right]
  \rhob_c \dot{\Zb} \; \delta Z
\label{simp_q}
\end{equation}
What remains is now to eliminate $\dot{\delta Z}$. This can be done using the perturbative version of eq (70) of \cite{Pourtsidou:2013aa}
which gives
\begin{eqnarray}
 \frac{\dot{\delta Z}}{\dot{\Zb}}
&=&
 \Psi 
+
\Bigg\{ 
  \left[\Zb \Kb_{YY}  +\rhob_c \frac{d}{dZ}\left(\frac{\beta_Z}{1+\Zb \beta}\right)\right] 
 \frac{1}{ \Kb_Y - \frac{\rhob_c \beta_Z}{1 + \Zb \beta}  }
 \nonumber
 \\
&&
  + \frac{ 3 \Kb_Y \adotoa + \Kb_{Y\phi} \dot{\phib}}{ 3 \Zb \Fb_Y  \adotoa  + a \Kb_\phi }
\Bigg\}\delta Z 
- \frac{3\Zb \Fb_Y}{ 3 \Zb \Fb_Y  \adotoa  + a \Kb_\phi  } (\dot{\Phi}+ \adotoa \Psi) 
 \nonumber
\\ 
&&
 +\Zb  \Bigg[
 \frac{3\Fb_{Y\phi} \adotoa \dot{\phib}-a^2 \Kb_{\phi\phi} - \Fb_Y \grad^2 }{  3 \Zb \Fb_Y  \adotoa  + a \Kb_\phi    }
\nonumber
\\
&&
+ \frac{ a \Kb_{Y\phi}}{ \Kb_Y - \frac{\rhob_c \beta_Z}{1 + \Zb \beta}   }
 \Bigg] \theta_\phi 
+ \frac{\rhob_c \beta_Z}{ (1+\Zb \beta  ) \Kb_Y - \rhob_c \beta_Z  }\delta_c
\label{eq:pertKGT2nonadiab}
\end{eqnarray}
Using (\ref{eq:pertKGT2nonadiab}) and (\ref{eq:deltaZgen}) into (\ref{simp_q}) we may now determine the  coefficients. 
They are
\begin{eqnarray}
&& 
A_1 = A_6 = 0   \; , \qquad 
A_2 =  \frac{\Zb \Fb_Y}{ 6 \Zb \Fb_Y  \adotoa  + 2a \Kb_\phi  } Q
\nonumber
\\
&&
A_3 =  \frac{c_s^2 }{(1+w)\left(  \Kb_Y - \frac{\rhob_c \beta_Z}{1 + \Zb \beta}  \right)}
\Bigg\{ 
Q \bigg[
 \Zb  \Kb_Y\frac{d}{dZ} \ln\left(\frac{\Zb\beta_Z}{1+ Z\beta}\right)
 \nonumber
 \\
&&
\ \ 
+ \Zb^2 \Kb_{YY} 
\bigg]
-  \frac{\rhob_c \beta_Z  \left[ Q + \Zb^2 \left(3 \Kb_Y \adotoa + \Kb_{Y\phi} \dot{\phib}  \right) \right]
}{1 + \Zb \beta}
\Bigg\}
\nonumber
\\
&&
  A_4 =  \frac{  (1+\Zb \beta  ) \Kb_Y}{ (1+\Zb \beta  ) \Kb_Y - \rhob_c \beta_Z  }  Q 
\nonumber
\\
&&
A_5 =
 \frac{1 }{ \Kb_Y - \frac{\rhob_c \beta_Z}{1 + \Zb \beta}  }
\Bigg\{
a \Kb_\phi Q 
\bigg[  \frac{\rhob_c \beta_Z}{\Zb \Kb_Y(1 + \Zb \beta)}  
 - \frac{\Zb \Kb_{YY}}{\Kb_Y}
 \nonumber
 \\
&&
 -  \frac{d}{dZ} \ln\left(\frac{\Zb\beta_Z}{1+ Z\beta}\right)
\bigg] 
 +  \frac{\Zb\rhob_c  \beta_Z ( 3\Kb_Y \adotoa + \Kb_{Y\phi} \dot{\phib} )  }{1+\Zb \beta} 
 \frac{a\Kb_\phi}{\Kb_Y}
 \nonumber
\\ 
&&
  - \frac{\Zb^2 \rhob_c \beta_Z  (3 \Fb_{Y\phi} \adotoa \dot{\phib}-a^2 \Kb_{\phi\phi} - \Fb_Y \grad^2  ) }{1+\Zb \beta}
 + Q\Zb  a \Kb_{Y\phi}
\Bigg\}
\nonumber
\\
 && 
B_5 = Q \qquad  B_{i\ne 5} = 0
\label{type_2_coeff}
\end{eqnarray}

%%%%%%%%%%%%%%%%%%%%%%%%%%%%%%%%%%%%%%%%%%%%%%%%%%%%%%%%%%%%%%%%%%%%%%%%%%%%%%%%%%%%%%%%

\subsection{Type 3 theory of DM coupled to DE}
\label{type_3_sec}
Type 3 models are classified via
\be
L(n,Y,Z,\phi)=F(Y,Z,\phi)+f(n).
\ee
The coupling current in this case is    \cite{Pourtsidou:2013aa}
\be
J_{\nu}=q^\beta_{\;\;\nu}  \bigg\{ X \nabla_\beta \phi  + F_Z  \nabla_\beta  Z + Z  F_Z    u^\mu \nabla_\mu u_\beta\bigg\}.
\ee
where $X \equiv \nabla_\mu(F_Z u^\mu)$
A straightforward calculation gives 
\be
 Q = q = 0
\ee 
(although there can be 2nd order corrections to $J_0$)
This means that the Type 3 case provides for a pure momentum-transfer coupling up-to linear order in perturbation theory.

To proceed to the coefficients we need $S$ which is found to be 
\be
S = -\left( \Xb  \dot{\phib} + \Fb_Z \dot{\Zb}    + \Zb  \Fb_Z  \adotoa  \right) \theta_c - \Zb  \Fb_Z   \dot{\theta}_c - \frac{1}{a} \Fb_Z \dot{\varphi}
+ \Xb   \varphi
\ee
where the background value of $X$ is 
\begin{equation}
\Xb  = \frac{1}{a} \left[ (\Zb \Fb_{ZY} - \Fb_{ZZ})\dot{\Zb} -  \Fb_{Z\phi} \dot{\phib} -3\adotoa \Fb_Z \right]
\end{equation}
We eliminate the  $\dot{\theta}_c$ term using (\ref{CDM_eq_CN}) to get
\be
S=\frac{1}{1-\frac{\Zb \Fb_Z}{\rhob_c}}
  \left[\Xb\varphi -(\Xb\dot{\phib}+\Fb_Z\dot{\Zb})\theta_c+ \Fb_Z \delta Z \right].
\label{eq:Sres}
\ee
where to remind the reader $\delta Z =  -\frac{1}{a} \left( \dot{\varphi}-\dot{\phib} \Psi\right)$.
Now we need to express $\varphi$ and $\dot{\varphi}$  in terms of the fluid variables. From  \cite{Pourtsidou:2013aa} we find
\begin{equation}
\rhob_{DE} = \Zb^2 F_Y - \Zb F_Z + F \qquad \Pb_{DE} = -F
\label{eq:rhobPbT3}
\end{equation}
for the background variables while
\begin{eqnarray}
\delta \rhob_{DE} &=& 
\Zb \left[ F_Y - \Zb^2 F_{YY} +2\Zb F_{YZ} - F_{ZZ} \right] \delta Z
\nonumber 
\\
&&
\ \ \ + \left[\Zb^2 F_{Y\phi} - \Zb F_{Z\phi} + F_\phi \right] \varphi  
\label{eq:deltarhophitype3}
\\
\delta \Pb_{DE} &=&  
 ( \Zb  F_Y -  F_Z) \delta Z
 -  F_\phi \varphi
\label{eq:deltapphitype3}
\\
 \theta_{DE} &=& \frac{\frac{F_Y}{a} \varphi + F_Z \theta_c }{ F_Z - \Zb F_Y }
\label{eq:drhophidpphi_type3}
\end{eqnarray}
for the perturbations.
For completeness, the adiabatic sound speed is
\begin{eqnarray}
 c_a^2 &=& 
 \frac{ 3  \adotoa  ( \Zb F_Y  - F_Z)  - a \left[ F_\phi +  \Zb^2 F_{Y\phi} - \Zb F_{Z\phi} \right]
}{3\adotoa \Zb (\Fb_Y  + 2 \Zb \Fb_{YZ} - \Zb^2 \Fb_{YY} - F_{ZZ})}
\nonumber
\\
&&
-\frac{aF_\phi}{3\adotoa(  \Zb \Fb_Y - \Fb_Z  )}
\end{eqnarray}
while  the effective sound speed $c_s^2$ is
\begin{equation}
c^2_s=\frac{\Zb\Fb_Y-\Fb_Z}{\Zb\left(\Fb_Y+2\Zb\Fb_{YZ}-\Fb_{ZZ}-\Zb^2\Fb_{YY}\right)}.
\end{equation}
and the relative entropy parameter is
\be
 \mu = \frac{3 F_Z}{\Zb \Fb_Y} (c_s^2 - c_a^2) (\rhob_{DE} + \Pb_{DE})\adotoa  
\ee

Clearly if $F_\phi =0$ then the perturbations are adiabatic, i.e. $c_s^2 = c_a^2$ and $\mu=0$ (so that $\Pi_{DE}  = c_a^2 \delta_{DE}$).

\begin{center}
\begin{table*}
\begin{tabular}{|c||c|c|c|c|c|c|c|c|c|c|c|c|c|}
\hline
 Model/Coefficients   & $Q$    &$A_1$&$A_2$&$A_3$&$A_4$ & $A_5$ & $A_6$ & $B_1$ & $B_2$ & $B_3$ & $B_4$ & $B_5$ & $B_6$
\\
\hline
\hline
 Coupled Quintessence & $-\beta_A \rhob_c \dot{\phib}$ & -   &  -  &$\frac{Q}{1+w}$ & $Q$ & $\beta_A \rhob_c a^2 V_\phi$   & -  &  - & -  & -  &  -  & $Q$  &  -
\\
\hline
 $J_\mu \propto u_\mu$ & $a \Gamma_{int} \rhob_c$ & -  & -  & -  & $Q$  & - & - & -  & - & -  & -  &  - &  $Q$
\\
\hline
 elastic scattering &  - & - & - & - & - & - & - & - & - & - & - &  $- \rhob_{DE}(1 + w)an_D\sigma_D$  & $-B_5$
\\
\hline
Type-1 & $-\rhob_c \alpha_\phi \dot{\phib}$  & - & -  & $\frac{Q c_s^2 }{1 + w}$ & $Q$ & $Q\left[\frac{\alpha_{\phi\phi}}{\alpha_\phi} - \frac{ c_s^2 \dot{\phib} \Kb_\phi}{(1+w) \Kb}\right]$  &-  &- &- & - & - & Q & - 
\\
\hline
Type-2 &  $\frac{\Zb \beta_Z\rhob_c}{1+\Zb \beta}\dot{\Zb}$ & - & $A_2$ & $A_3$ &  $A_4$  & $A_5$  & - &  - &  - & - &  - & $Q$ &  -
\\
\hline
Type-3 & - & -  & -  & - & -  & -  & - & -  & - & $B_3$ & - &  $B_5$ & $-B_5 +  \frac{ 3  \adotoa\Zb F_Z c_s^2  }{1-\frac{\Zb \Fb_Z}{\rhob_c}} $
\\
\hline
\end{tabular}
\caption{Specific models and their PPF coefficients. The coupled Quintessence model is a subcase of Type 1 with $\alpha_\phi = \beta_A$. The elastic scattering model is in fact distinct from Type-3 (see text at the end of section \ref{type_3_sec}).
For the coefficients $A_2$, $A_3$, $A_4$ and $A_5$ in the case of Type-2 see (\ref{type_2_coeff}). For the coefficients $B_3$ and $A_5$ in the case of Type-3 see (\ref{type_3_coeff}).  For the remaining functions the
reader is referred to each specific example in the text.
}
\label{tab_of_models}
\end{table*}
\end{center}

We can now proceed to find the coefficients. Equations (\ref{eq:deltarhophitype3}) and (\ref{eq:drhophidpphi_type3}) can be inverted to give
\begin{eqnarray}
 \delta Z &=& 
 \left[ \frac{\mu}{\Fb_Z} - \frac{aF_\phi}{F_Y} \right]  \left[ \theta_{DE} +  \frac{\Zb F_Z}{\rhob_{DE}+\Pb_{DE}} \theta_c \right]
\nonumber 
\\
&&
+ \frac{c_s^2\Zb}{1+w}  \delta_{DE} 
\end{eqnarray}
and
\begin{eqnarray}
 \varphi = a \left( \frac{  \Fb_Z }{\Fb_Y} - \Zb \right) \theta_{DE} -\frac{ a\Fb_Z }{\Fb_Y} \theta_c
\end{eqnarray}
We also need the equation for $\dot{\Zb}$ which is found to be (eq (75) in  \cite{Pourtsidou:2013aa} )
\begin{eqnarray*}
 \dot{\Zb}  =- 3\adotoa  \Zb \left[ c_a^2  + \frac{aF_\phi}{3\adotoa(  \Zb \Fb_Y - \Fb_Z  )} \right]
\end{eqnarray*}
The above equations are then inserted into (\ref{eq:Sres}) to give the required coefficients as
\begin{eqnarray}
B_1 &=& B_2 = B_4 = 0
\nonumber
\\
B_3  &=& \frac{1}{1-\frac{\Zb \Fb_Z}{\rhob_c}}  \frac{\Zb \Fb_Z c_s^2}{1+w}
\nonumber
\\
B_5 &=&
 \frac{a}{1-\frac{\Zb \Fb_Z}{\rhob_c}}
  \bigg[
 \Xb \left( \frac{  \Fb_Z }{\Fb_Y} - \Zb \right)
+  \Fb_Z   \left[ \frac{\mu}{a\Fb_Z} - \frac{F_\phi}{F_Y} \right] 
\bigg]
\nonumber
\\
B_6 &=& - B_5  + \frac{ 3  \adotoa\Zb F_Z c_s^2  }{1-\frac{\Zb \Fb_Z}{\rhob_c}} 
\label{type_3_coeff}
\end{eqnarray}

It would seem tempting to try model the elastic scattering model~\cite{Simpson:2010vh} discussed above (section \ref{elastic_scattering_model}) into the Type 3 class. 
However, this is in fact impossible. As we can easily check, the
elastic scattering model requires $B_3= 0$. Within the  Type 3 class this is possible only if $F$ is independint of $Z$ (i.e. $F_Z=0$). This implies that
$B_5$ and $B_6$ are also zero, in other words, the model becomes completely uncoupled. Hence, it is impossible construct a model of elastic scattering between CDM and DE
 within the Type-3 class of coupled Dark Energy.

\section{Conclusions}
\label{sec:conclusions}

We have presented the most general parameterization of models of Dark Energy which is explicitly coupled to Dark Matter using the Parameterized Post-Friedmannian framework, and have shown that 
it is able to encapsulate a rich variety of theories. 

Starting from the linearised Einstein equations and using the Bianchi identities we managed to express the modifications to GR coming from the dark sector coupling as a collection of new terms containing the metric potentials and their derivatives as well as the scalar modes of the two dark sector components, i.e. the fluid variables of (generalised) Dark Matter and Dark Energy. Of course, our formalism is based on a few basic assumptions: the background cosmology has an FRW solution, all field equations are at most second-order in time derivatives, and the field equations are gauge-invariant. Completing the parameterization we were left with 24 free functions, but demanding gauge invariance we derived 4 constraint equations which eliminated 4 free functions. 

Twenty free functions in our general parameterization is certainly a big number, but by imposing certain well motivated assumptions, for instance that
 the Dark Matter is Cold, that the Dark Energy is shear-less and that the pressure perturbation is not a dynamical quantity we reduced the number of free functions to 12.
Furthermore, we showed that only a handful of these functions are non-zero when one considers known models. We demonstrated this by investigating a number of specific models in the literature, as well as the classes of theories we constructed in \cite{Pourtsidou:2013aa}. It is useful to note that, although our theories in \cite{Pourtsidou:2013aa} are derived from an action, the PPF parameterization does \emph{not} require knowledge of the action, but only knowledge of the field equations. This means that the PPF parameterization is a very useful tool for phenomenological model building (see \cite{Baker:2012zs} for further discussion in the context of modified gravity theories).  The full list of models we consider in this work is displayed in table \ref{tab_of_models} along with their coefficients.

Our Type 1, 2 and 3 theories contain a fairly general coupling function and hence they encapsulate many different models. The parameterization coefficients for these theories can depend, of course, on the chosen coupling function and its derivatives, and other quantities such as the background coupling $Q$, the background field energy density $\rhob_\phi$, the quintessence potential $V(\phi)$, the speed of sound $c^2_s$ etc. For Type 1 theories there is only 1 non-zero $B$ coefficient and 3 non-zero $A$ coefficients, for Type 2 there is 1 non-zero $B$ coefficient and 4 non-zero $A$ coefficients, while for Type 3 all $A$'s are automatically zero and there are 3 non-zero $B$'s: different classes of theories correspond to different non-zero functions.  In particular, from all the cases we have studied, the coefficients
$A_1$, $A_6$, $B_1$, $B_2$ and $B_4$ are always zero. It would be indeed very interesting to find models  for which any of these coefficients is non-zero.

It would also be interesting to consider the inverse problem, i.e. given $Q(t)$ and a set of PPF coefficients $A_i$ and $B_i$ can we reconstruct the functions that appear in the coupled Dark Energy Lagrangian for each of the Type-1,2 and 3 theories? For instance, what kind of functions correspond to constant PPF coefficients? Tackling the inverse problem will help to reduce the free functions into simple functional forms which are parametrised by a set of constants and can make constraining such theories easier and more efficient.

Another important question is how we can further constrain the PPF functions?
As discussed in \cite{Baker:2012zs}, we might expect to find that a subset of the PPF functions can be very well constrained, whilst another subset can not.
 However this might not pose a serious problem, as the constraining power of a few PPF functions might be sufficient to distinguish between theories.  Tackling the inverse problem will certainly help here as it can guide us to which functional form of the PPF coefficients is the most useful.
The implementation of the PPF framework presented here in numerical codes for the computation of the cosmological effects of interacting Dark Energy could provide an answer to these questions. 
We plan to investigate this in future work.

\section{Acknowledgments}
E.C. acknowledges support from STFC grant ST/L0003934/1.
A.P. acknowledges support from STFC grant ST/H002774/1. 
A.P. acknowledges the University of Nottingham for hospitality during various stages of this work. 
C.S. acknowledges initial support from the Royal Society and further support from the European Research Council.
The research leading to these results has received funding from the European Research Council
under the European Union's Seventh Framework Programme (FP7/2007-2013) / ERC Grant Agreement n. 617656 ``Theories
 and Models of the Dark Sector: Dark Matter, Dark Energy and Gravity''. 
\bibliographystyle{apsrev}
\bibliography{references}

\end{document}